\documentclass[aps,pre,twocolumn,superscriptaddress,showpacs]{revtex4-1}
\usepackage{amsfonts}
\usepackage{epsfig}
\usepackage{amsmath,amssymb}
\usepackage{times}
\usepackage{xcolor}
\usepackage{multirow}

\begin{document}

\title{Synergistic interactions promote behavior spreading and alter phase transition on multiplex networks}

\date{\today}

\author{Quan-Hui Liu}
%\email{tangminghuang521@hotmail.com}
\affiliation{Web Sciences Center, School of Computer Science and
  Engineering, University of Electronic Science and
Technology of China, Chengdu 611731, China}
\affiliation{Big Data Research Center, University of Electronic Science and Technology of China, Chengdu 611731, China}
\affiliation{ Laboratory for the Modeling of Biological and
  Socio-technical Systems, Northeastern University, Boston, MA 02115,
  USA}

\author{Wei Wang}
\affiliation{Web Sciences Center, School of Computer Science and
  Engineering, University of Electronic Science and
Technology of China, Chengdu 611731, China}
\affiliation{Big Data Research Center, University of Electronic Science and Technology of China, Chengdu 611731, China}
\affiliation{College of Computer Science and Technology, Chongqing University of Posts and
Telecommunications, Chongqing 400065, China}

\author{Shi-Min Cai}
\affiliation{Web Sciences Center, School of Computer Science and
  Engineering, University of Electronic Science and
Technology of China, Chengdu 611731, China}
\affiliation{Big Data Research Center, University of Electronic Science and Technology of China, Chengdu 611731, China}
\affiliation{Center for Polymer Studies and Department of Physics, Boston University, Boston, MA 02215, USA}

\author{Ming Tang}
\email{tangminghan007@gmail.com}
\affiliation{Web Sciences Center, School of Computer Science and
  Engineering, University of Electronic Science and
Technology of China, Chengdu 611731, China}
\affiliation{Big Data Research Center, University of Electronic Science and Technology of China, Chengdu 611731, China}
\affiliation{School of Information Science Technology, East China Normal University, Shanghai 200241, China}

\author{Ying-Cheng Lai}
\affiliation{School of Electrical, Computer and Energy Engineering, Arizona State University, Tempe, Arizona 85287, USA}

\begin{abstract}

Synergistic interactions are ubiquitous in the real world. Recent studies 
have revealed that, for a single-layer network, synergy can enhance spreading 
and even induce an explosive contagion. There is at the present a growing 
interest in behavior spreading dynamics on multiplex networks. What is the 
role of synergistic interactions in behavior spreading in such networked
systems? To address this question, we articulate a synergistic behavior 
spreading model on a double layer network, where the key manifestation of the 
synergistic interactions is that the adoption of one behavior by a node in 
one layer enhances its probability of adopting the behavior in the other 
layer. A general result is that synergistic interactions can greatly enhance 
the spreading of the behaviors in both layers. A remarkable phenomenon is 
that the interactions can alter the nature of the phase transition associated 
with behavior adoption or spreading dynamics. In particular, depending on the 
transmission rate of one behavior in a network layer, synergistic interactions 
can lead to a discontinuous (first-order) or a continuous (second-order) 
transition in the adoption scope of the other behavior with respect to its 
transmission rate. A surprising two-stage spreading process can arise: due to 
synergy, nodes having adopted one behavior in one layer adopt the other 
behavior in the other layer and then prompt the remaining nodes in this layer 
to quickly adopt the behavior. Analytically, we develop an edge-based 
compartmental theory and perform a bifurcation analysis to fully understand, 
in the weak synergistic interaction regime where the dynamical correlation 
between the network layers is negligible, the role of the interactions in 
promoting the social behavioral spreading dynamics in the whole system.

\end{abstract}
%\pacs{89.75.Hc,87.19.X-,87.23.Ge}

\maketitle

\section{Introduction} \label{sec:intro}
A central problem in network science and engineering is to understand,
predict, and control the dynamics of virus or information spreading on
complex networks~\cite{BBV:book,vespignani2012modelling,PRCVV:2015}. Social 
contagion processes such as the propagation of an opinion, diffusion of a 
belief, and spread of a particular behavior, occur commonly in the real 
world~\cite{valente:1996,CF:2007,Young:2009,rogers:2010,centola:2011,
banerjee:2013,Zha2016,zhang2016dynamics}.
With the modern technological advances, a variety of online social networking
platforms (e.g., \emph{Facebook} and \emph{Youtube}) have become a
routine necessity for a substantial fraction of individuals in the entire
population. Spreading dynamics in modern online social networks have
attracted a great deal of recent attention and a variety of mathematical
models have been articulated to understand and predict the relevant
phenomena~\cite{PRCVV:2015,CFL:2009,DW:2004,DW:2005}.
For example, the threshold model, a binary state spreading model, was
introduced earlier to address the phenomenon of behavior adoption, where
a node in a social network adopts a new behavior only when the
number~\cite{Granovetter:1973} or the fraction~\cite{Watts:2002} of its
nearest adopted neighbors exceeds a threshold value. A representative
threshold model reveals the phenomenon that the final size of the nodes
adopting the behavior first grows continuously and then decreases
discontinuously as the mean degree of the network is 
increased~\cite{Watts:2002}. Within the threshold model, the effects of
parameters and network structure on the dynamics of social behavioral
spreading have been studied, which include the initial seed 
size~\cite{GC:2007}, the clustering 
coefficient~\cite{whitney:2010,hackett:2011,zhuang:2017}, the
community structure~\cite{glesson:2008,nematzadeh:2014} and
multiplexity~\cite{brummitt:2012,yaugan:2012,cozzo:2013}. The dynamical 
process described by the threshold model, however, is Markovian because the 
state of a node depends only on the current state of its neighbors. The 
original model is thus not able to encompass an important aspect of real 
contagion dynamics: social reinforcement originated from the memory
effect~\cite{dodds:2004,krapivsky2011reinforcement,ZLZ:2013,LWTZ:2016} - a
feature that is characteristically non-Markovian. To overcome this deficiency
of the classical threshold model, a non-Markovian behavior spreading model
taking into account the received cumulative pieces of behavioral information
for any node to adopt the behavior was introduced~\cite{WTZL:2015}. A
prediction of the modified model is that the dependence of the final
behavior adoption size on the information transmission rate can change
from being discontinuous to being continuous through continuous changes
in the dynamical or structural parameters. The non-Markovian behavior
spreading model also allows additional issues such as the heterogeneity of
adoption thresholds~\cite{WTSW:2016}, the limited contact
capacity~\cite{WSZTZ:2015}, and the effect of temporal network
structure~\cite{liu2016social} to be addressed.

Most previous works on network behavior spreading focused on a single 
social behavior contagion process through empirical 
methods~\cite{centola:2011,banerjee:2013} and mathematical 
models~\cite{DW:2004,DW:2005,Granovetter:1973,Watts:2002,WTZL:2015,LWTZL:2017}.
In the real world, it is common for
two or more distinct behaviors to spread simultaneously in a social system,
where interactions between the corresponding spreading processes inevitably
arise. For example, individuals who have adopted \emph{Windows} services are
more likely to use other services from the same company, e.g.
\emph{Microsoft Office}. In online networking systems, two different tweets
on the same event or subject can diffuse on the twitter network at the same
time. The user seeing one tweet will experience an increased exposure
to the other tweet, and vice versa, since these two tweets are closely
related. In this case, the two tweets spread synergistically as they
mutually prompt each other in the process of retweeting~\cite{myers:2012}.
The synergistic mechanism is also typical in the adoption of online services.
A good example is the adoption of two online services, say \emph{Google}
and \emph{Youtube} through two types of tweets: one containing the URLs
with \emph{google} and another with \emph{youtube}. The numbers of the two
types of tweets are synchronized most of the time, implying that they are
synergistic to each other~\cite{zarezade:2015}. The synergistic effect also
occurs in disease spreading, where the interaction between pathogens may
mutually strengthen their spreading process, and such an effect may have
played a role in the co-epidemic of the Spanish flu and pneumonia in
1918~\cite{taubenberger20061918,brundage2008deaths,Chen:2013,cai2015avalanche,
hebert2015complex}. In spite of its ubiquity, the synergistic mechanism among 
two or more simultaneously spreading behaviors was not investigated in previous
studies~\cite{DW:2004,DW:2005,Granovetter:1973,Watts:2002,WTZL:2015}.

In this paper, we articulate a synergistic social behaviors spreading model
to address and understand the impacts of synergistic interactions among
multiple behaviors on their spreading. As the spreading of each behavior 
typically occurs on a different network layer, it is necessary to incorporate 
a multilayer network structure~\cite{DGPA:2016,KABGP:2014,BBCR:2014}. To be 
concrete, we consider the spreading dynamics of two distinct behaviors in 
two-layer coupled networks, where each layer supports the spreading of one 
behavior with its own transmission path, as described by a non-Markovian 
process. The synergistic mechanism between the two behavior adoption dynamics
is that, once a node adopts a behavior in one layer, it becomes more
susceptible to adopting the other behavior that spreads in the other
network layer. We develop an edge-based compartmental theory to analyze
and understand how the synergistic interactions impact the simultaneous
spreading dynamics of the behaviors. We find, as suggested by intuition,
that the synergistic interactions greatly facilitate the adoption of both
behaviors. However, surprisingly, a phenomenon is that the adoption of one
behavior can lead to a characteristic change in the adoption of the other
behavior: its final adoption size versus its information rate can change
from being discontinuous to continuous, where the former corresponds to a
first-order phase transition while the latter to a second-order transition.
Remarkably, the synergistic effect can induce a two-stage contagion process,
in which nodes having adopted one behavior in one layer will adopt the other
behavior in the other layer. When there is a sufficient number of seeds, i.e.,
when the number of nodes having adopted the other behavior in the other
layer is sufficiently large, the remaining nodes will adopt the behavior
quickly. While it is intuitively understandable that the synergistic
interactions can promote the spreading dynamics of the distinct behaviors
involved, our work lays a quantitative foundation for this phenomenon. Our
model will not only serve as a useful framework to understand the interplay
between synergy and simultaneous spreading of multiple behaviors or diseases,
but will also provide insights into predicting or even controlling the
underlying dynamics. Due to the ubiquity of synergy in different fields
such as social science, computer science, biology and biomedicine, broad
relevance of our model is warranted.

In Sec.~\ref{sec:model}, we describe the network and the synergistic 
behavior spreading models. In Sec.~\ref{sec:theory}, we carry out a detailed 
theoretical analysis. In Sec.~\ref{sec:numerics}, we present extensive 
simulation results with respect to the theoretical predictions. In 
Sec.~\ref{sec:discussion}, we summarize the main results and discuss a few 
pertinent issues. 

\section{Model} \label{sec:model}

There are two components in our model: multiplex networks and spreading
dynamics of synergistic behaviors. We first introduce the model of multiplex 
networks, and then present the synergistic behavior spreading model.

\subsection{Model of multiplex networks}

In general, network layers in an interdependent networked system have
different internal structures and dynamical functions. To capture the
essential dynamics of simultaneous spreading of distinct behaviors, we focus 
on multiplex networks~\cite{DGPA:2016,KABGP:2014,BBCR:2014}. Consider the
simple setting of a duplex system consisting of two layers or subnetworks.
Initially, we generate two independent layers, denoted as $a$ and $b$,
which have the same node set and support the spread of behaviors $1$ and
$2$, respectively. We use the configuration model~\cite{CBP:2005}
to generate each subnetwork, where the degree distribution $P_a(k_a)$ of
layer $a$ is completely independent of the distribution $P_b(k_b)$ of layer
$b$. For large and sparse subnetworks, the configuration model stipulates
that both interlayer and intralayer degree-degree correlations are negligible.

\subsection{Synergistic behavior spreading model}

We use a representative non-Markovian spreading model, the
susceptible-adopted-recovered (SAR)~\cite{WTZL:2015} model, to describe
the dynamics of behavior spreading, and then introduce the synergistic
mechanism between the spreading processes of the two behaviors.

For each behavior $c\in\{1,2\}$, at any time a node will be in one of the
three states: susceptible ($S_c$), adopted ($A_c$) and recovered ($R_c$).
A node in state $S_c$ has not adopted behavior $c$ but it has an interest
in $c$. A node in the $A_c$ state has adopted the behavior and can transmit
the information about the behavior (denoted as information $c$) to its
neighbors. The node loses interest in transmitting the information when it
is in the $R_c$ state. The evolution process of behavior $c$ can be
described, as follows. Initially, $\rho_c(0)$ fraction of nodes are randomly
chosen as the nodes that have adopted the behavior and the remaining nodes
are set to be in the susceptible state. At each time step,
each node in the $A_c$ state transmits the information to each of its
susceptible neighbors with the transmission rate $\lambda_c$. Suppose a
neighboring node $v$ already has accumulated $m-1$ pieces of information
$c$ from its distinct neighbors. One more successful transmission will
make the number of information pieces to become $m$. We assume
non-redundant information transmission, i.e., once an adopted node has
transmitted the information to node $v$, the former will not transmit
the same information to latter again. If the cumulative number $m$ pieces
of information $c$ that the susceptible node $v$ has is equal to or larger
than a threshold, the node will adopt the behavior $c$ and changes its state
to $A_c$. Simultaneously, each $A_c$ node will turn to the $R_c$ state
at the recovery rate $\gamma_c$. The behavior spreading process will
terminate when all the adopted nodes have recovered. More specifically,
$\rho_1(0)$ and $\rho_2(0)$ are the fractions of nodes randomly chosen
as seeds (i.e., adopted nodes) for behavior $1$ and $2$ on each layer,
respectively, where the remaining nodes are in the susceptible state.
Information $1$ ($2$) diffuses in layer $a$ ($b$) with transmission rate
$\lambda_1$ ($\lambda_2$), and the recovery rates for behaviors $1$ and $2$
are $\gamma_1$ and $\gamma_2$, respectively.

In the general SAR model, each susceptible node has its own adoption threshold
for a behavior. However, for simplicity in modeling the synergistic interaction
between the spreading of the two behaviors, we assume that all nodes have the
same adoption threshold for each behavior: we denote the adoption threshold
for behavior 1 in layer $a$ as $T_1$ and that for behavior 2 in layer $b$ as
$T_2$. As a manifestation of mutual synergy, a node having adopted one behavior
will become more susceptible to adopting the other behavior. To quantify
the synergistic effect, we assume that, once node $i$ has adopted behavior
$1$ ($2$), it will generate an increase ${\Delta}T_2>0$ (${\Delta}T_1>0$) in
the number of pieces of information about behavior $2$ ($1$). The quantities
${\Delta}T_1$ and ${\Delta}T_2$ thus characterize the strength of the
synergistic effect, and we have ${\Delta}T_1\in[0, T_1]$ and
$\Delta{T_2}\in[0,T_2]$. For ${\Delta}T_1 = 0$, a node having adopted
behavior $2$ in layer $b$ will not impact on its adoption of behavior $1$ in
layer $a$. Similarly, the adoption of behavior $1$ will have no effect on
adopting behavior $2$ if ${\Delta}T_2=0$. If a node has adopted behavior $2$,
it will adopt behavior $1$ only if ${\Delta}T_1+m{\geq}T_1$, where $m$
represents the number of cumulative pieces of behavioral information $1$
in layer $a$ that this node has received from distinct neighbors.

\section{Theory} \label{sec:theory}

We exploit the edge-based compartmental 
theory~\cite{WTZL:2015, miller2012edge,yang2012epidemic,karrer:2010} to 
analyze the dynamical process of behavior spreading subject to synergistic 
interactions, under the assumption that each subnetwork is large and sparse 
with no internal degree-degree correlations. We also assume that the degree
distribution of network $a$ is completely independent of that of
network $b$, so interlayer degree-degree correlation can be neglected too.
The fraction of nodes in each state can be treated as a continuous
variable. For each behavior $c\in\{1,2\}$, we denote $S_c(t)$, $A_c(t)$
and $R_c(t)$ as the fractions of nodes being in the susceptible, adopted,
and recovered state, respectively, for behavior $c$ in the corresponding
layer at time $t$. During the spreading process, the susceptible nodes 
adopting behavior $c$ decreases the value of $S_c(t)$ but leads to an 
increase in $A_c(t)$, and the recovery of the adopted nodes for behavior $c$ 
decreases $A_c(t)$ but increases $R_c(t)$. Using these notations, the 
dynamical evolution equations for behavior $c$ can be written as
\begin{equation} \label{eq:e1}
\frac{dA_{c}(t)}{dt}=-\frac{d{S_c}(t)}{dt}-\gamma_c{A_c(t)}
\end{equation}
and
\begin{equation} \label{eq:e2}
\frac{d{R_c}(t)}{dt}={\gamma_c}A_c(t).
\end{equation}
For $t\rightarrow\infty$, the states of all individuals remain unchanged
and $R_c(\infty)$ is the final adoption fraction of behavior $c$.

\subsection{Edge-based compartmental theory} \label{subsec:ebc_theory}

Despite that the spreading processes of behaviors 1 and 2 occur in
different networks ($a$ and $b$, respectively) and the dynamical parameters
such as the information transmission rates ($\lambda_1$ and $\lambda_2$),
the recovery rates ($\gamma_1$ and $\gamma_2$), and the adoption thresholds
($T_1$ and $T_2$), are different, the mathematical equations governing the
underlying processes have identical forms. It thus suffices to derive the
equations for behavior $1$ spreading in layer $a$.

To solve Eqs.~(\ref{eq:e1}) and~(\ref{eq:e2}), we need to calculate the
fraction of susceptible nodes for behavior $1$ at time step $t$. Firstly,
for nodes of degree $k_a$ in layer $a$, two cases can arise where the nodes
do not adopt behavior 1: (1) these nodes have not adopted behavior $2$ on
layer $b$ and the cumulative number of received pieces of information $1$
in layer $a$ is less than $T_1$, and (2) these nodes have already adopted
behavior $2$ in layer $b$, but the cumulative number of received pieces of
information $1$ in layer $a$ is less than $T_1-{\Delta}T_1$. Under the
assumption that there is no dynamical correlation between the layers, we have
that the fraction of susceptible nodes of degree $k_a$ for behavior $1$ at
time $t$ is given by
\begin{eqnarray} \label{eq:e3}
S_1(k_a,t)&=&
\nonumber
S_2(t)\sum_{m=0}^{T_1-1}\phi_1(k_a,m,t)\\
&+&[1-S_2(t)]\sum_{m=0}^{T_1-1-{\Delta}T_1}\phi_1(k_a,m,t).
\end{eqnarray}
In Eq.~(\ref{eq:e3}), the first term on the right side is the probability
that a node of degree $k_a$ in layer $a$ at time $t$ does not adopt behavior 1.
This term contains two parts that describe the following two situations,
respectively: (1) the received cumulative number of pieces of information 1
is less than $T_1$ with probability $\sum_{m=0}^{T_1-1}\phi_1(k_a,m,t)$, and
(2) with probability $S_2(t)$, a random node in layer $b$ does not adopt
behavior 2 at time $t$ (i.e., a node in layer $b$ does not adopt behavior 2
and is still in the susceptible state), where the quantity $\phi_1(k_a,m,t)$
is the probability for a node of degree $k_a$ to have received $m$ pieces of
information $1$ by time $t$ in layer $a$. Combining the two parts, we find
that the first term is identical to the second term in Eq.~(\ref{eq:e3}).
Using the degree distribution of network $a$, we can express the fraction
of susceptible nodes for behavior $1$ as
\begin{eqnarray} \label{eq:e4}
S_1(t)=\sum_{k_a}P_{a}(k_a)S_1(k_a,t).
\end{eqnarray}
In Eq.~(\ref{eq:e3}), the quantity $\phi_1(k_a,m,t)$ can be expressed as
\begin{eqnarray} \label{eq:e5}
\phi_1(k_a,m,t) = [1-\rho_1(0)]B_{k_a,m}[\theta_1(t)],
\end{eqnarray}
where $B_{k,m}(w)$ denotes the binomial distribution
$B_{k, m}(1-w)^mw^{k-m}$ and $\theta_{1}(t)$ is the probability
that a random neighbor $v$ of node $u$ in layer $a$ has not transmitted the
behavioral information $1$ to node $u$ by time $t$. To take into account the
dynamical correlations among the states of the adjacent nodes, we make use of
the cavity theory~\cite{WTZL:2015, miller2012edge,yang2012epidemic,karrer:2010}
to analyze the quantity $\theta_{1}(t)$, where node $u$ is in the cavity state 
so that it cannot transmit the behavioral information to its neighbors but it 
can receive the information from its neighbors.

To solve Eqs.~(\ref{eq:e3}) and~(\ref{eq:e4}), we need the value of
$\theta_1(t)$ [the computation of $S_2(t)$ is the same as that of $S_1(t)$].
Noting that a random neighbor $v$ of node $u$ in layer $a$ can be in one of
the following three states: $S_1$, $A_1$ and $R_1$, we have that
$\theta_{1}(t)$ is the sum of the probabilities that the neighbor $v$ does
not transmit information $1$ to $u$ when $v$ is in the $S_1$, $A_1$ or $R_1$
state. We have
\begin{eqnarray}\label{eq:e6}
\theta_{1}(t)=\xi_{1}^{S}(t)+\xi_{1}^{A}(t)+\xi_{1}^{R}(t),
\end{eqnarray}
where $\xi_{1}^{S}(t)$ [$\xi_{1}^{A}(t)$ or $\xi_{1}^{R}(t)$] denotes the
susceptible (adopted or recovered) neighbor $v$ of $u$ which has not
transmitted information $1$ to node $u$ up to time $t$ in layer $a$.

\begin{figure*}
\centering
\includegraphics[width=\linewidth]{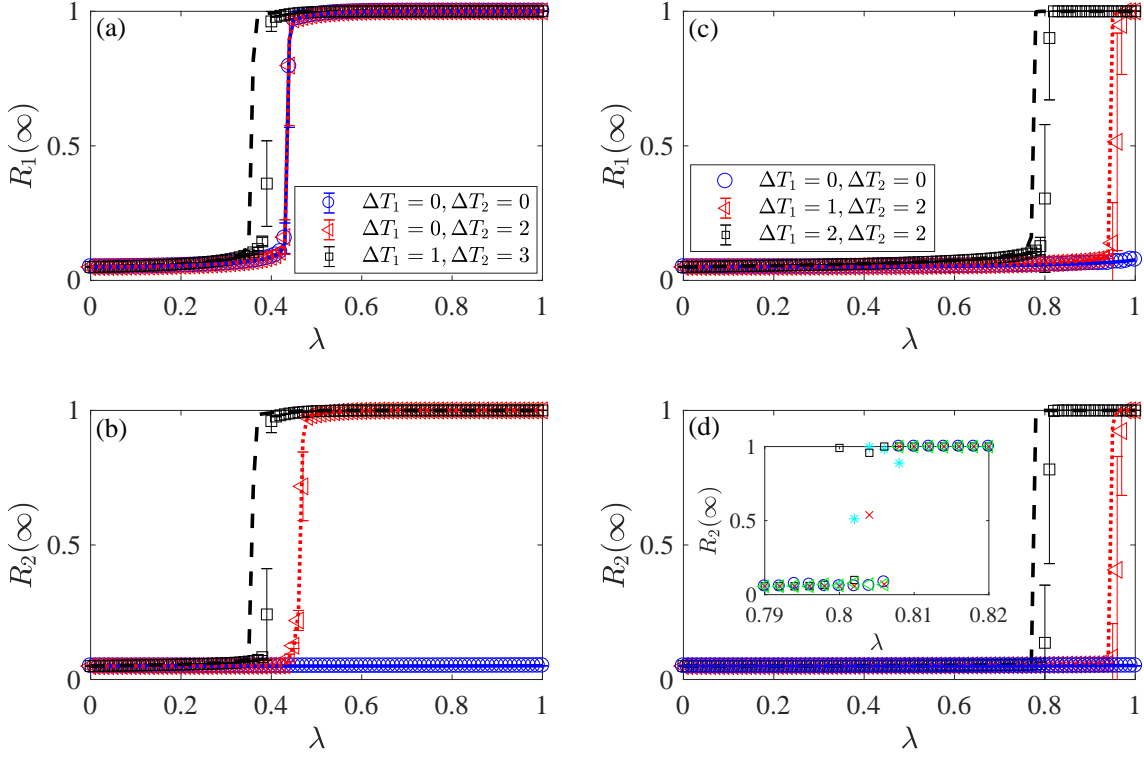}
\caption{ \bf Effect of synergistic strength on behavior spreading for random
regular double-layer networks (RR-RR)}. (a,b) For $T_1=2$ and $T_2=4$, the fractions
$R_1(\infty)$ and $R_2(\infty)$ of recovered nodes in layers $a$ and $b$,
respectively, versus $\lambda$, where $\lambda_1=\lambda_2=\lambda$. 
(c,d) The corresponding plots for a different set of threshold values:
$T_1=3$ and $T_2=4$. The symbols are direct simulation results while the
lines are the corresponding theoretical prediction obtained by iterating
Eqs.~(\ref{eq:e1})-(\ref{eq:e4}) and (\ref{eq:e14})-(\ref{eq:e17}). The plots in the inset of (d) are results from 
five stochastic simulations for the parameter settings 
($\Delta T_1=2$, $\Delta T_2=2$, $T_1=3$ and $T_2=4$). The network sizes of both layers are set as $N=5*10^4$, the simulation results are average by using 20 multiplex network realizations and each multiplex network is with $10^3$ independent dynamical realizations. Other parameters are $\gamma_1=\gamma_2=1$.
\label{fig:RR_RR}
%\end{center}
\end{figure*}

Suppose a random neighbor $v$ of degree $k_a^{\prime}$ of node $u$ is
susceptible initially, node $u$ cannot transmit information $1$ to $v$
since $u$ is in the cavity state. Node $v$ can only receive the information
from its other $k_a^{\prime}-1$ neighbors. The probability that node $v$ has
received $m$ pieces of information $1$ in layer $a$ by time $t$ is then
\begin{eqnarray}\label{eq:e7}
\tau_1(k_a^{\prime},m,t)=B_{k_a^{\prime}-1,m}[\theta_1(t)].
%\binom{k_a^{\prime}-1}{m}{[\theta_1(t)]}^{k_{a}^{'}-m-1}{[1-\theta_{1}(t)]}^{m}.
\end{eqnarray}
Similar to Eq.~(\ref{eq:e3}), we have that the probability that the neighboring
node $v$ is still in the susceptible state for behavior $1$ at time $t$ is 
given by
\begin{eqnarray} \label{eq:e8}
\Phi_{1}[k_a^{\prime},\theta_1(t),\theta_2(t)] &=&
S_2(t)\sum_{m=0}^{T_1-1}\tau_1(k_a^{\prime},m,t) \\ \nonumber
&+&[1-S_2(t)]\sum_{m=0}^{T_1-1-{\Delta}T_1}\tau_1(k_a^{\prime},m,t).
\end{eqnarray}
For uncorrelated networks, the probability for a random edge to connect a node
of degree $k_a^{\prime}$ is $k_a^{\prime}P(k_a^{\prime})/\langle k_a \rangle$,
where $\langle k_a \rangle$ is the average degree of network layer $a$.
A neighboring node in the susceptible state cannot transmit the behavioral
information. Thus, $\xi_{1}^{S}(t)$ is equal to the probability that the
neighboring node is in the susceptible state, which is
\begin{equation} \label{eq:e9}
\xi_{1}^{S}(t)=[1-\rho_1(0)]\frac{\sum_{k_a^{\prime}}k_a^{\prime}
P(k_a^{\prime})\Phi_{1}[k_a^{\prime},\theta_1(t),\theta_2(t)]}
{\langle k_a \rangle}.
\end{equation}
If a random neighbor $v$ is in the adopted state for behavior 1, success
in information transmission from node $v$ to node $u$ will result in a
decrease in $\theta_{1}(t)$. We thus have
\begin{equation} \label{eq:e10}
\frac{d{\theta_{1}(t)}}{dt}=-{\lambda_1}\xi_{1}^{A}(t).
\end{equation}
At the same time, once the adopted neighbor $v$ has recovered before it
can transmit information $1$ to node $u$, there will be an increase in
$\xi_{1}^{R}(t)$. (Note that here we use the synchronous updating rule,
meaning that the transmission and recovery events happen consecutively in
discrete time steps.) The increase in $\xi_{1}^{R}(t)$ contains two parts
that describe the following two situations, respectively: (1) with probability
$1-\lambda_1$, the neighboring node $v$ has not transmitted information 1
to $u$, and (2) simultaneously, node $v$ recovers with probability $\gamma_1$.
Combining these two parts, we obtain the increment of $\xi_{1}^{R}(t)$ as
\begin{equation} \label{eq:e11}
\frac{d\xi_{1}^{R}(t)}{dt}={\gamma_1}(1-\lambda_1)\xi_{1}^{A}(t).
\end{equation}
Combining Eqs.~(\ref{eq:e10}) and (\ref{eq:e11}), we obtain an explicit
expression for $\xi_{1}^{R}(t)$:
\begin{equation} \label{eq:e12}
\xi_{1}^{R}(t)=\frac{\gamma_1[1-\theta_{1}(t)](1-\lambda_1)}{\lambda_1}.
\end{equation}
Inserting Eqs.~(\ref{eq:e9}) and (\ref{eq:e12}) into Eq.~(\ref{eq:e6}),
we can write $\xi_{1}^{A}(t)$ as
\begin{eqnarray} \label{eq:e13}
\xi_{1}^{A}(t)&=&
\nonumber
\theta_{1}(t)-\frac{\sum_{k_a}^{'}k_a^{\prime}P(k_a^{\prime})
\Phi_{1}[k_a^{\prime},\theta_1(t),\theta_2(t)]}{\langle k_a \rangle}\\
&-&\frac{{\gamma_1}[1-\theta_{1}(t)](1-\lambda_1)}{\lambda_1}.
\end{eqnarray}
Substituting Eq.~(\ref{eq:e13}) into Eq.~(\ref{eq:e10}), we get the time
evolution of $\theta_{1}(t)$ as
\begin{eqnarray} \label{eq:e14}
\frac{d{\theta_1(t)}}{dt}&=&
\nonumber
-{\lambda_1}\theta_1(t)+ \gamma_1[1-\theta_1(t)](1-\lambda_1)\\
\nonumber
&+&\lambda_1(1-\rho_1(0))\\
&\times&\frac{\sum_{k_a^{\prime}}k_a^{\prime}P(k_a^{\prime})\Phi_{1}[k_a^{\prime},\theta_1(t),\theta_2(t)]}{\langle k_a \rangle}.
\end{eqnarray}
Following a similar procedure, we can derive the expression of $\theta_2(t)$,
the probability that a random neighbor $v$ of node $u$ in layer $b$
has not transmitted the behavioral information $2$ to node $u$ by time $t$,
and $S_2(k_b,t)$. We have
\begin{eqnarray} \label{eq:e15}
\frac{d{\theta_2(t)}}{dt}&=&
\nonumber
-{\lambda_2}\theta_2(t)+ \gamma_2[1-\theta_2(t)](1-\lambda_2)\\
\nonumber
&+&\lambda_2(1-\rho_2(0))\\
&\times&\frac{\sum_{k_b^{\prime}}k_b^{\prime}P(k_b^{\prime})
\Phi_{2}[k_b^{\prime},\theta_1(t),\theta_2(t)]}{\langle k_a \rangle}
\end{eqnarray}
and
\begin{eqnarray} \label{eq:e16}
S_2(k_b,t)&=&
\nonumber
S_1(t)\sum_{m=0}^{T_2-1}\phi_2(k_b,m,t)\\
&+&[1-S_1(t)]\sum_{m=0}^{T_2-1-{\Delta}T_2}\phi_2(k_b,m,t),
\end{eqnarray}
where the form of $\Phi_{2}[k_b^{\prime},\theta_1(t),\theta_2(t)]$ in
Eq.~(\ref{eq:e15}) is similar to
$\Phi_{1}[k_a^{\prime},\theta_1(t),\theta_2(t)]$, and $\phi_2(k_b,t)$
in Eq.~(\ref{eq:e16}) is similar to $\phi_1(k_a,t)$. It is thus not
necessary to write down the expressions again. Using the degree distribution
of network $b$, we have the fraction of susceptible nodes at time $t$ in
layer $b$ as
\begin{equation} \label{eq:e17}
S_2(t) =\sum_{k_b}P_{b}(k_b)S_2(k_b,t).
\end{equation}
Iterating Eqs.~(\ref{eq:e1})-(\ref{eq:e4}) and~(\ref{eq:e14})-(\ref{eq:e17}),
we can obtain the fractions of susceptible nodes at time $t$ in both layers:
$S_1(t)$ and $S_2(t)$. In addition, we can substitute $S_1(t)$ [$S_2(t)$] into
Eqs.~(\ref{eq:e1}) and (\ref{eq:e2}) and calculate the fractions of the adopted
nodes and of the recovered nodes in layer $a$ ($b$) at time $t$. Taking
the limit $t\rightarrow\infty$, we can obtain the final fractions of adoption
of the two behaviors. Results on the final adoption fractions from direct
numerical simulations together with the corresponding theoretical predictions
for different parameter values are shown in Fig.~\ref{fig:RR_RR}. We obtain
a good agreement between theory and numerics. For example, for $T_1=2$ and
$T_2=4$, Fig.~\ref{fig:RR_RR}(b) shows that, without the synergistic effect
of behavior 1, i.e., $\Delta{T_2}=0$, behavior 2 will not exhibit any
outbreak. For $\Delta{T_2}=2$, behavior 2 is adopted globally. When there
are mutual synergistic effects, e.g., $\Delta{T_1}=1$ and $\Delta{T_2}=3$
or $T_1=3$ and $T_2=4$, the adoption of both behaviors is enhanced, as
shown in Figs.~\ref{fig:RR_RR}(c) and \ref{fig:RR_RR}(d), respectively. Note that
there are some outliers (e.g., there are one black square in Fig. 1 (a) and two black squares in Fig. 1 (d)) around the critical transmission rate since the SAR model is not a deterministic threshold model, which is in contrast to the Watts threshold model. The randomness exists in the process of simulations when the behavior information transmission rate is smaller than 1. Supposing a susceptible node with adoption threshold equal to 3, when it has three adopted neighbors it will not adopt the behavior if one of its adopted neighbor does not succeed in transmitting the behavior information. As shown in the inset of Fig. 1(d), there are some stochastic simulations that $R_2(\infty)$ does not increase from a very smaller value to a value close 1 directly. 

A fundamental issue in spreading dynamics in complex networks is phase
transitions~\cite{PRCVV:2015}. As a system parameter (e.g., the infection
rate) changes through a critical point, the final size of the infected nodes
starts to increase from zero. An abrupt and discontinuous increase in the
final size signifies a first-order phase transition, while a gradual and
continuous change is indicative of a second-order phase transition. An
objective of our study is then to uncover and understand the effect of
synergistic interactions on the phase transitions associated with the social
behavior spreading dynamics. To analyze the phase transition, we focus on
the fixed point (root) of Eqs.~(\ref{eq:e14}) and (\ref{eq:e15}) associated
with the final state (i.e., $t\rightarrow \infty$). Simplifying notation
as $\theta_1 \equiv \theta_1(\infty)$ and $\theta_2 \equiv \theta_2(\infty)$,
we write Eqs.~(\ref{eq:e14}) and (\ref{eq:e15}) as
\begin{equation} \label{eq:e18}
\theta_1=f_1(\theta_1,\theta_2),
\end{equation}
and
\begin{equation} \label{eq:e19}
\theta_2=f_2(\theta_1,\theta_2),
\end{equation}
respectively, where
\begin{eqnarray} \label{eq:e20}
f_1(\theta_1,\theta_2)&=& \nonumber
\frac{[1-\rho_1(0)]\sum_{k_a^{\prime}}k_a^{\prime}P_a(k_a^{\prime})
\Phi_{1}(k_a^{\prime},\theta_1,\theta_2)}{\langle k_a \rangle}\\
&+&\frac{\gamma_1}{\lambda_1}[1-\theta_1](1-\lambda_1),
\end{eqnarray}
and
\begin{eqnarray} \label{eq:e21}
f_2(\theta_1,\theta_2) &=& \nonumber
\frac{[1-\rho_2(0)]\sum_{k_b^{\prime}}k_b^{\prime}P_b(k_b^{\prime})
\Phi_{2}(k_b^{\prime},\theta_1,\theta_2)}{\langle k_b \rangle}\\
&+&\frac{\gamma_2}{\lambda_2}[1-\theta_2](1-\lambda_2).
\end{eqnarray}
Because of the nonlinear functions $\Phi_1(k_a^{\prime},\theta_1,\theta_2)$ 
in Eq.~(\ref{eq:e20}) and $\Phi_2(k_a^{\prime},\theta_1,\theta_2)$ in
Eq.~(\ref{eq:e21}), to analyze the whole parameter space is infeasible.
We thus focus on some representative or benchmark cases to gain certain
analytic understanding of the numerical results. Specifically, we consider
two cases in terms of the adoption thresholds of the two behaviors:
(1) the adoption threshold of one behavior is less than that of the
other behavior ($T_1<T_2$ or $T_1>T_2$), and (2) $T_1=T_2$.

\subsection{Solutions for $T_1<T_2$} \label{subsec:T1ltT2}

\begin{figure*}
\centering
\includegraphics[width=\linewidth]{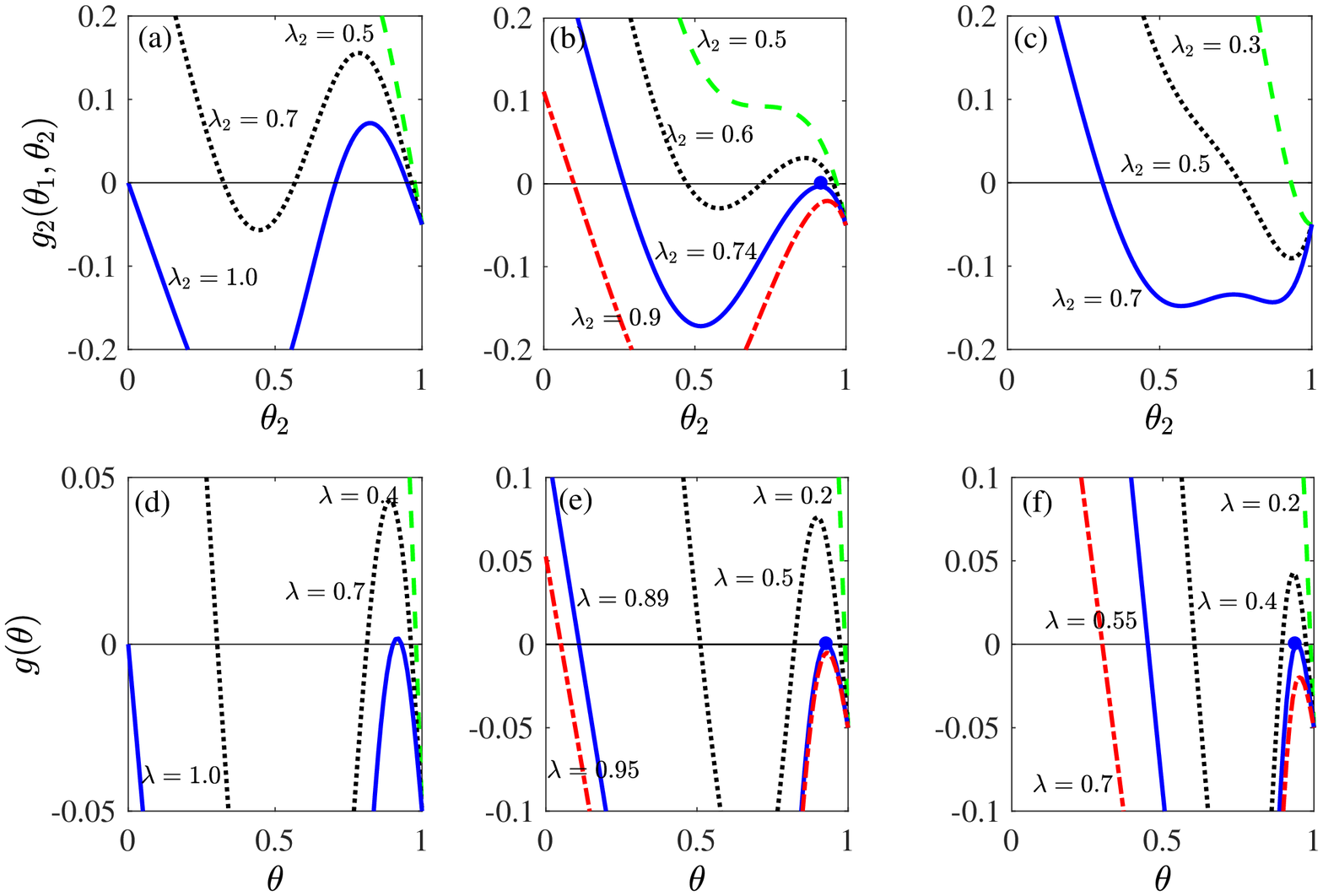}
\caption{ {\bf Phase transitions associated with simultaneous behavioral
spreading on double-layer random regular networks.} The graphical
solutions of Eqs.~(\ref{eq:e18}), (\ref{eq:e19}) and (\ref{eq:e24}) are presented. The upper
panels show the results for the case $T_1<T_2$, i.e., $T_1=1$ and $T_2=4$,
where $g_2(\theta_1,\theta_2)$ is plotted as a function of $\theta_2$ for
$\Delta{T_2}=0$ (a), $\Delta{T_2}=2$ (b) and $\Delta{T_2}=3$ (c). The
fixed points of Eqs.~(\ref{eq:e17}) and (\ref{eq:e18}) are the intersections
between the respective curves and the horizontal axis. Other parameters are
${\Delta}T_1=0$, $\lambda_1=0.12$, and $\rho_1(0)=\rho_2(0)=0.05$. The
lower panels show the cases of $T_1=T_2$ for $T_1=T_2=3$,
${\Delta}T_1={\Delta}T_2=\Delta{T}$, and $\lambda_1=\lambda_2=\lambda$,
where the values of $g(\theta)$ are plotted as a function of $\theta$ for
$\Delta{T}=0$ (d), $\Delta{T}=1$ (e) and $\Delta{T}=2$ (f). The fixed points
of Eq.~(\ref{eq:e24}) are the intersections between the respective curves
and the horizontal axis. The initial adoption fraction is
$\rho(0)=0.05$. The blue dots in (b), (e) and (f) denote the points of
tangency. Other parameters are $\gamma_1=\gamma_2=1$.}
\label{fig:bifurcation}
\end{figure*}

For $T_1<T_2$, ${\Delta}T_1=0$ and ${\Delta}T_2>0$, indicating
that the adoption of behavior $2$ has no effect on the spread of behavior $1$
but the adoption of the latter will enhance the spread of former, as shown
in Fig.~\ref{fig:RR_RR}. Because Eqs.~(\ref{eq:e18}) and (\ref{eq:e19}) are nonlinear functions of 
$\theta_1$ and $\theta_2$, typically there are multiple roots. In addition, 
there is persistent transmission of behavioral information from individuals in 
an adopted state (i.e., $A_1$ or $A_2$) to their neighbors, so $\theta_1(t)$
and $\theta_2(t)$ decrease with time. Thus, if Eqs.~(\ref{eq:e18}) and 
(\ref{eq:e19}) possess more than one stable fixed point, only the one with 
the maximum value is physically meaningful~\cite{WTZL:2015}.
Since Eq.~(\ref{eq:e18}) contains the parameters $\lambda_1$ and $\theta_1$ 
only, for a given value of $\lambda_1$, we can obtain the value of $\theta_1$. 
For given values of the parameters $\lambda_2$ and ${\Delta}T_2$,
with $\theta_1$ we can solve Eq.~(\ref{eq:e19}) numerically. As shown in top
panel of Fig.~\ref{fig:bifurcation}, we see that Eq.~(\ref{eq:e19}) typically
has a non-zero trivial solution even for small values of $\lambda_2$,
indicating that, even when the initial adopted fraction of behavior $2$ is
small (e.g., $\rho_2(0)=0.05$), it will always be adopted by a certain fraction
of the nodes. However, the initial fraction of seeds will have an effect
on the final adoption size~\cite{GC:2007,WTZL:2015}. To better focus on
the effect of synergistic interactions on simultaneous spreading of the
two behaviors, we set $\rho_1(0) = \rho_2(0)=0.05$ and calculate the
final adoption size versus the behavioral information transmission rate
with a particular eye on the possible type of phase transitions.

For $\Delta{T_2}=0$, the number of roots (fixed points) of the function
$g_2(\theta_1,\theta_2)=f_2(\theta_1,\theta_2)-\theta_2$ is 1 or 3,
as shown in Fig.~\ref{fig:bifurcation}(a). Because the physically meaningful
solution is the maximum value of the stable fixed point of Eq.~(\ref{eq:e19}),
there is no global outbreak in behavior $2$ [verified numerically, see
Fig.~\ref{fig:deltaEffect}(a)]. For ${\Delta}T_2=2$, the function
$g_2(\theta_1,\theta_2)$ is tangent to the horizontal axis at $\theta_2^c$
for the critical value of $\lambda_2^c\approx 0.74$. Further increasing
$\lambda_2$ above $\lambda_2^c$ removes the tangent point and leaves
$g_2(\theta_1,\theta_2)$ with only one intersection point with the horizontal
axis. Importantly, from the standpoint of bifurcation analysis, we see that,
at this point, the physically meaningful fixed point $\theta_2$ decreases
abruptly to a small value, signifying a first-order phase transition. The
critical value $\lambda_2^c$ for a given $\lambda_1$ can be obtained by using
the criterion that a nontrivial solution of Eq.~(\ref{eq:e19}) emerges, which
corresponds to the point at which the function $g_2(\theta_1,\theta_2)$ is
tangent to horizontal axis at the critical value of $\theta_2^c$. That is,
the critical condition for this case can be obtained by combining
Eqs.~(\ref{eq:e18}) and (\ref{eq:e19}) and the following equation
\begin{equation} \label{eq:e22}
\frac{dg_2(\theta_1,\theta_2)}{d\theta_2}|_{\theta_2^c}=0.
\end{equation}
For ${\Delta}T_2=3$ and $\lambda_1=0.12$, Eq.~(\ref{eq:e19}) has a
single root whose value decreases with $\lambda_2$, as shown in
Fig.~\ref{fig:bifurcation}(c). This means that $R_2(\infty)$ increases with
$\lambda_2$ continuously.

For a given value of the transmission rate $\lambda_1$ of behavior $1$, the
critical condition is then that behavior $2$ will be adopted if its
transmission rate $\lambda_2$ is larger than $\lambda_2^c$. Similarly, we can
compute the minimal information transmission rate of behavior 1 required for
a global outbreak of behavior $2$. In particular, setting $\lambda_2$ to
be the maximum value (i.e., $\lambda_2=1.0$) and substituting it into
Eqs.~(\ref{eq:e19}) and (\ref{eq:e22}), we get the critical values of
$\theta_1$ and $\theta_2$. Substitute these values into Eq.~(\ref{eq:e18}),
we obtain $\lambda_1^{m}$, the minimal information transmission rate of
behavior $1$.

Numerical solutions of Eq.~(\ref{eq:e19}) also show that, for large values
of $\lambda_1$ and ${\Delta}T_2>0$, it has one fixed point only when
varying $\lambda_2$, so $R_2(\infty)$ increases with $\lambda_2$ continuously.
As a result, there exists the critical parameter value $\theta_1^c$
(i.e., $\lambda_1^{c}$), across which the dependence of $R_2(\infty)$ on
$\lambda_2$ changes from being discontinuous to continuous. For the special
case of $T_1<T_2$ (e.g., $T_1=1$, $T_2=4$, ${\Delta}T_1=0$ and
${\Delta}T_2>0$), we can numerically solve Eqs.~(\ref{eq:e19}) and
(\ref{eq:e22}), together with the condition~\cite{BDGM:2010}
\begin{equation}\label{eq:e23}
\frac{d^2g_2(\theta_1,\theta_2)}{d\theta_2^2}|_{\theta_2^c}=0.
\end{equation}
Once $\theta_1^c$ is determined, we can substitute the value of $\theta_1^c$
into Eq.~(\ref{eq:e18}) to get $\lambda_1^{c}$. In particular, $R_2(\infty)$
increases with $\lambda_2$ discontinuously for $\lambda_1<\lambda_1^{c}$ and
the increasing pattern becomes continuous for $\lambda_1\ge\lambda_1^{c}$.
Using the same approach, we can determine the critical value of $\lambda_2^c$
above (below) which $R_2(\infty)$ increases with $\lambda_1$ discontinuously
(continuously).

\subsection{Solutions for $T_1=T_2$}

This is the symmetric case where ${\Delta}T_1={\Delta}T_2=\Delta{T}$,
$\lambda_1=\lambda_2=\lambda$, $\langle k_a \rangle=\langle k_b \rangle$,
and $P_a(k)=P_b(k)=P(k)$. The symmetry implies $\theta_1(t)=\theta_2(t)$ and
$f_1(\theta_1,\theta_2)=f_2(\theta_1,\theta_2)$. For simplicity, we denote
$\theta(t)\equiv \theta_1(t)$ and
$f[\theta(t)]\equiv f_1[\theta_1(t),\theta_2(t)]$.
Equations~(\ref{eq:e18})-(\ref{eq:e21}) can be written as
\begin{equation} \label{eq:e24}
\theta=f(\theta),
\end{equation}
where
\begin{eqnarray} \label{eq:e25}
\nonumber
f(\theta)&=&\frac{[1-\rho(0)]\sum_k{kP(k)\Phi(k,\theta)}}{\langle k \rangle}
+\frac{\gamma}{\lambda}(1-\theta)(1-\lambda).
\end{eqnarray}
Similar to treating Eq.~(\ref{eq:e8}), we have
\begin{eqnarray} \label{eq:e26}
\Phi(k,\theta) &=& S(\infty)\sum_{m=0}^{T-1}B_{k-1,m}(\theta) \\ \nonumber
&+& [1-S(\infty)]\sum_{m=0}^{T-1-\Delta{T}}B_{k-1,m}(\theta).
\end{eqnarray}
The final fraction of the susceptible nodes of behavior $1$ ($2$) in
layer $a$ ($b$) is given by
\begin{eqnarray} \label{eq:e27}
\nonumber
S(\infty)&=&[1-\rho(0)]\sum_k{P(k)\{S(\infty)\sum_{m=0}^{T-1}B_{k,m}(\theta)}\\
&+&{[1-S(\infty)]\sum_{m=0}^{T-1-\Delta{T}}B_{k,m}(\theta)\}}.
\end{eqnarray}
Using the same analysis method as for the case $T_1<T_2$, we find that the
number of fixed points of Eq.~(\ref{eq:e24}) is $1$ or $3$, as shown in
the lower panel of Fig.~\ref{fig:bifurcation}. Whether there is a tangent point
between the function $g(\theta)=f(\theta)-\theta$ and the horizon axis
depends on the strength $\Delta{T}$ of synergistic interactions. For
$\Delta{T}=0$, there is no tangent point and only the maximum value of
the fixed point of Eq.~(\ref{eq:e24}) is physically meaningful, indicating
that behavior $2$ is adopted by a small fraction of nodes only. For
$\Delta{T}=1$ and $\Delta{T}=2$, the function $g(\theta)$ can be tangent
to the horizon axis, as shown in Figs.~\ref{fig:bifurcation}(e) and
\ref{fig:bifurcation}(f). When $\lambda_2$ is increased passing through
$\lambda_2^c$, the tangent point disappears and the function $g(\theta)$
has only one intersecting point with the horizontal axis. In this case,
the fixed point $\theta$ changes discontinuously to a small value, signifying
a first-order phase transition.

\begin{figure*}
\centering
\includegraphics[width=\linewidth]{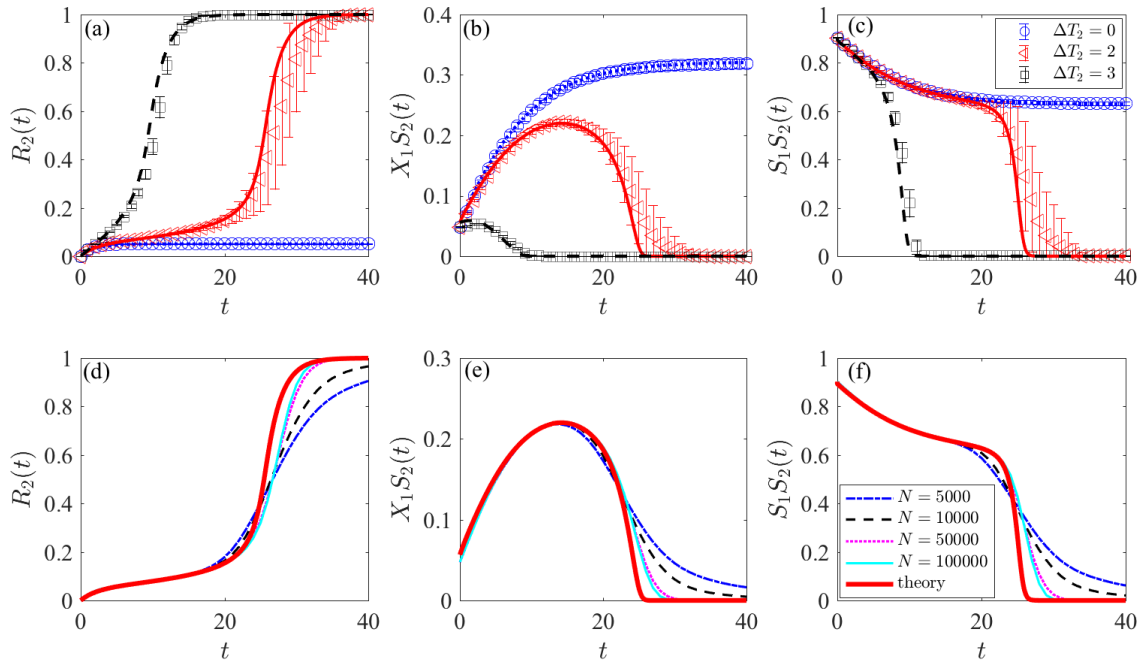}
\caption{ {\bf Time evolution of behavior spreading subject to synergistic
interactions}. For random regular double-layer networks, (a, d) the fraction
of recovered nodes $R_2(t)$ versus time $t$, (b, e) the fraction of nodes
in state $X$ in layer $a$ and in state $S$ in layer $b$ versus time,
(c, f) the fraction of nodes in the $S$ state in both layers $a$ and
$b$ versus time. (d)-(f) are the simulation results when $\Delta T_2=2$ for different network sizes $N$. The parameters are $\lambda_1=0.06$,
$\lambda_2=0.8$, $T_1=1$, $T_2=4$, and ${\Delta}T_1=0$.
The symbols are simulation results and the lines are theoretical
prediction in (a)-(c). In the theoretical analysis of the state $X_1S_2(t)$,
dynamical correlations between the layers are ignored. Other parameters
are $\gamma_1=\gamma_2=0.5$. }
\label{fig:time_evolution}
\end{figure*}

\section{Numerical validation} \label{sec:numerics}

In this section, we perform extensive simulations of behavior spreading 
on different multiplex networks. We use the notation ``RR-RR" to denote 
the case where both layer $a$ and layer $b$ host the random regular 
networks. The notation ``ER-SF'' represents the setting where layer $a$ 
is an Erd\"{o}s-R\'{e}nyi (ER) random network~\cite{ER:1959} and layer 
$b$ hosts an scale-free (SF) network~\cite{BA:1999}. Other possible 
combinations are ``ER-ER'', ``SF-SF'' and ``SF-ER''. The size of each 
network is $N_a=N_b=5\times10^4$ and the average degree is 
$\langle k \rangle=10$ for both networks. The initial adoption fractions 
of behavior $1$ in layer $a$ and behavior $2$ in layer $b$ are set to be 
$\rho_1(0) = \rho_2(0)=0.05$. To calculate the pertinent statistical averages, we use $20$ multiplex 
network realizations and at least $10^3$ independent dynamical realizations 
for each parameter setting. Unless otherwise specified, the above parameters are adopted in the simulations.
Let $X_1$ denote the situation where a node is in the $A$ or $R$ state in
layer $a$ so, for example, the notion $X_1S_2$ means that, in layer $a$, a
node is in the adopted state or recovered state but it is in the susceptible
state in layer $b$. Similarly, $A_1S_2$ indicates that a node is in the
adopted state in layer $a$ and is in the susceptible state in layer $b$, which
means that the node adopts behavior $1$ but not behavior 2.

\subsection{RR-RR multiplex networks}

We first perform direct numerical simulations of behavioral spreading dynamics
on double layer networked systems consisting of two random regular networks
to provide support for our theoretical predictions.

Our theoretical analysis in Sec.~\ref{subsec:T1ltT2} gives that, for
$T_1<T_2$, synergistic interactions can promote behavior adoption and
spreading. To be concrete, we set $T_1=1$ and $T_2=4$.
Figure~\ref{fig:time_evolution}(a) shows the time evolution of the fraction
$R_2(t)$ of the recovered nodes in layer $b$ for different values of the
synergistic interaction strength ${\Delta}T_2$. We see that behavior $2$
will not outbreak if ${\Delta}T_2=0$. For ${\Delta}T_2=2$ and ${\Delta}T_2=3$,
$R_2(t)$ exhibits a two-stage contagion process, where nodes having adopted
behavior $1$ in layer $a$ will first adopt behavior $2$, until when there
is a sufficient number of seeds (i.e., nodes having adopted behavior $2$)
in layer $b$ to stimulate the remaining nodes. When this happens, behavior $2$
will be adopted quickly in layer $b$. This phenomenon can be explained by
noting that, for a small fraction of the initial seeds for behavior 2 [i.e.,
$\rho_2(0)=0.05$], if the synergistic effect of adoption of behavior 1 is
absent [i.e., $\Delta{T_2}=0$], behavior 2 will not be adopted globally and
only the recovery of the seeds can lead to an increase in the value of
$R_2(t)$. Note that the number of $X_1S_2(t)$ nodes increases with the
adoption of behavior $1$ in layer $a$ [Fig.~\ref{fig:time_evolution}(b)]
since the $S_1$ nodes will change to $X_1$ nodes and there is no decrease
in the number of $S_2$ nodes in the network. For ${\Delta}T_2=2$, nodes that
have adopted behavior 1 are more likely to adopt behavior 2 as compared with
those that have not adopted behavior 1. Nodes having adopted behavior 1 in
layer $a$ will first adopt behavior 2 in layer $b$, as indicated by the
decrease in the number of the $X_1S_1(t)$ nodes in
Fig.~\ref{fig:time_evolution}(c). Before most of the $X_1S_2$ nodes have
adopted behavior 2, the seeds (i.e., adopted nodes for behavior 2) in layer
$b$ are sufficient to stimulate the remaining nodes to adopt behavior 2,
inducing a two-stage contagion process. A similar phenomenon occurs for
$\Delta{T_2}=3$. When the simulation results are compared with the theoretical predictions, we find the former matches well with the latter for $\Delta T_2=0$. While the deviation emerges when $\Delta T_2=2$, which are derived from the finite-size effects of the networks and the dynamical correlation between layers. From the bottom panels of Fig. 3, we will find the deviation is decreased when increasing the network size, but the deviation will still exist since the interlayer dynamical correlations are ignored in the theoretical method.

\begin{figure}
\centering
\includegraphics[width=\linewidth]{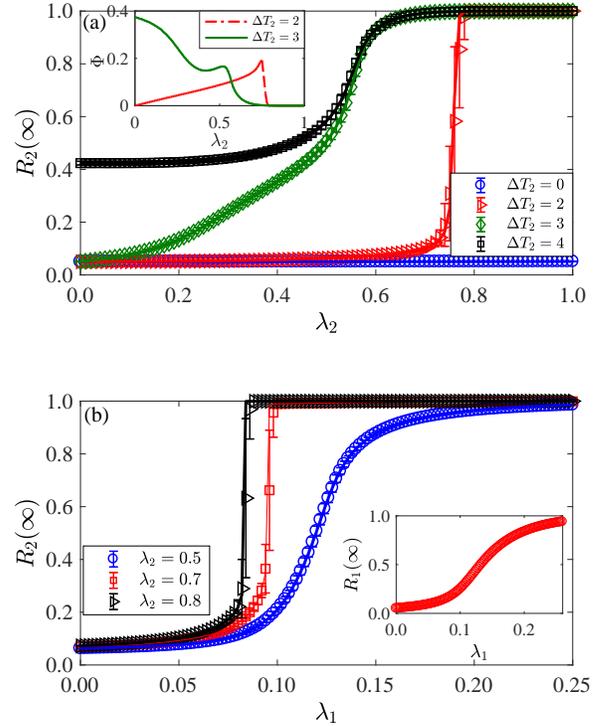}
\caption{ {\bf Asymptotic and stable adoption of behavior 2}.
For random regular double-layer networks, the final adoption size of
behavior 2 versus the information transmission rates: (a) $R_2(\infty)$
versus $\lambda_2$ for different values of the synergistic strength
$\Delta{T_2}$, where the transmission rate for behavior $1$ is
$\lambda_1=0.12$ and the corresponding fraction of the nodes adopting
behavior $1$ is $R_1(\infty)\approx0.393$, (b) $R_2(\infty)$ versus
$\lambda_1$ for different values of $\lambda_2$. The inset in (a) shows the final fraction $\Phi$ of nodes in the subcritical state for behavior 2. The subcritical state is defined as the state for a node that it will adopt the behavior when it receives one additional piece of information. The inset in (b) shows
the final adoption fraction of behavior $1$ in layer $a$ versus $\lambda_1$,
where ${\Delta}T_2=3$. The symbols are simulation results and the
lines (i.e., dotted, dotted dashed and solid lines) are theoretical prediction. Other parameters are $T_1=1$, $T_2=4$, ${\Delta}T_1=0$, and $\gamma_1=\gamma_2=1$.}
\label{fig:deltaEffect}
\end{figure}

Figure~\ref{fig:deltaEffect}(a) shows, for $T_1=1$, $T_2=4$ and
$\lambda_1=0.12$, $R_2(\infty)$ versus $\lambda_2$ for different values
of ${\Delta}T_2$, where the fraction of the $X_1S_2$ nodes in the system
is about $0.393$. As the synergistic interaction strength ${\Delta}T_2$
is increased, behavior $2$ is adopted more readily since the number of
information pieces about it is decreased. A remarkable phenomenon is
the characteristic change in the dependence of $R_2(\infty)$ on $\lambda_2$.
In particular, for ${\Delta}T_2=2$, $R_2(\infty)$ increases with $\lambda_2$
discontinuously but the increasing pattern becomes continuous for
${\Delta}T_2=3$. The reason for the characteristic change is that, for
${\Delta}T_2=2$, the nodes having adopted behavior $1$ still need to receive
additional \emph{two} (i.e., $T_2-{\Delta}T_2$) pieces of information to
adopt behavior $2$. The system will accumulate a relatively large number of nodes in the 
subcritical state when the behavioral information transmission rate approaches the critical point, as shown in the inset of Fig. 4(a). Therein, the subcritical state is defined as the node in such state will adopt the behavior if it receives one additional piece of behavior information~\cite{WTZL:2015}.
A slight increase in $\lambda_2$ will cause a node in this state to receive
an additional piece of information and thus adopts behavior $2$. The node
can then transmit the information to its neighbors, which will cause its
subcritical neighbors to adopt behavior $2$ accordingly, and so on,
leading to an avalanche of behavior adoption for the $X_1S_2$ nodes.
When most of the $X_1S_2$ nodes have adopted behavior 2 in an abrupt fashion,
there is a sufficient number of $A_2$ nodes in layer $b$ to stimulate the
remaining $S_1S_2$ nodes to adopt behavior 2. As a result, increasing
$\lambda_2$ slightly can lead to a discontinuous change in the value of
$R_2(\infty)$. However, for ${\Delta}T_2=3$, only one additional piece of information about 
behavior 2 is needed for the $X_1S_2$ nodes to adopt this behavior. As the
value of $\lambda_2$ is increased from zero, some $X_1S_2$ nodes may receive 
one piece of information about behavior 2 and adopt it, leading to a  
continuous decrease in the number of nodes in the subcritical state, as shown in the inset of Fig. 4 (b). This is equivalent to the dynamical process in the susceptible-infected-recovered (SIR) model, in 
contrast to the cascading process in, for example, the Watts threshold model. 
As a result, the value of $R_2(\infty)$ first increases with $\lambda_2$ 
continuously. When most of $X_1S_2$ nodes have adopted behavior 2, the fraction 
of adopted nodes in layer $b$ is sufficient to stimulate the remaining $S_1S_2$ nodes 
to adopt behavior 2. Since the fraction of adopted nodes is relatively large [e.g., 
$X_1(\infty)\approx0.393$], the value of $R_2(\infty)$ increases with 
$\lambda_2$ continuously~\cite{WTZL:2015} at a faster rate, as shown in 
Fig.~\ref{fig:deltaEffect}(a).
The same process occurs for $\Delta{T_2=4}$. These numerical results agree 
well with our bifurcation analysis based theoretical prediction.

Figure~\ref{fig:deltaEffect}(b) shows the dependence of $R_2(\infty)$ on
$\lambda_1$ for different values of $\lambda_2$. For a relatively small
value of $\lambda_2$ (e.g., $\lambda_2=0.5$), $R_2(\infty)$ increases
with $\lambda_1$ continuously, which can be understood by noting that,
in this case, a global adoption of behavior 2 requires more seeds in layer
$b$, and the spread of this behavior depends strongly on the spread of
behavior 1. However, for relatively large values of $\lambda_2$ (e.g.,
$\lambda_2=0.7$ and $\lambda_2=0.8$), $R_2(\infty)$ versus $\lambda_1$
can exhibit an abrupt or discontinuous increase. In this case, a slight
increase in the fraction of seeds for behavior 2 is sufficient for it
to spread globally by its own dynamics. Both the continuous growth for
small values of $\lambda_2$ and the discontinuous increase for larger
values of $\lambda_2$ are predicted by our bifurcation analysis based on
Eqs.~(\ref{eq:e18}), (\ref{eq:e19}), (\ref{eq:e22}) and (\ref{eq:e23})
by replacing $\theta_2$ with $\theta_1$ in Eqs.~(\ref{eq:e22}) and
(\ref{eq:e23}). There is a good agreement between numerics and theory.

\begin{figure}
\centering
\includegraphics[width=\linewidth]{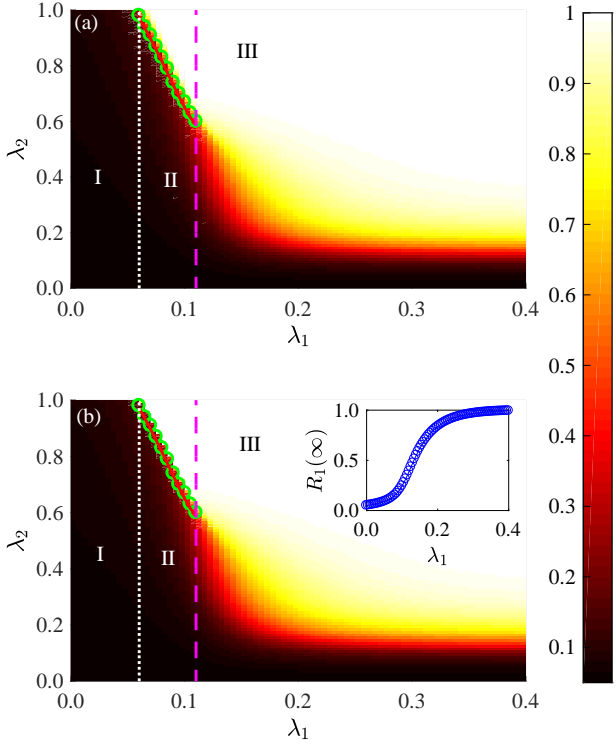}
\caption{ {\bf Dependence of final adoption size of behavior 2 on the
transmission rates}. For random regular networks, color coded values of
$R_2(\infty)$ in the parameter plane ($\lambda_1$, $\lambda_2$) of the two
information transmission rates: (a) numerical results and (b) theoretical
prediction based on solutions of Eqs.~(\ref{eq:e1})-(\ref{eq:e4}) and
(\ref{eq:e16})-(\ref{eq:e19}). The plane is divided into three regions by
the two vertical lines, where the dotted vertical line 
($\lambda_1=\lambda_1^m$) is from Eqs.~(\ref{eq:e18}), (\ref{eq:e19}) and 
(\ref{eq:e22}) for $\lambda_2=1$, and the dashed vertical line 
($\lambda_1=\lambda_1^c$) is determined by Eqs.~(\ref{eq:e18}), 
(\ref{eq:e19}), (\ref{eq:e22}) and (\ref{eq:e23}).
In region I, only a small fraction of the nodes is exposed to adopting
behavior $2$. In regions II and III, there are a discontinuous (first-order)
and a continuous (second-order) phase transition, respectively. The green
circles and the red line in region II, respectively, indicate the numerically
obtained critical information transmission rate of behavior 2 and the
theoretical prediction from Eqs.~(\ref{eq:e17}), (\ref{eq:e19}) and
(\ref{eq:e22}) for a given value of $\lambda_1$. The inset in (b) shows the
final adoption fraction of behavior $1$ versus the information transmission
rate of this behavior. Other parameters are $T_1=1$, ${\Delta}T_1=0$, $T_2=4$,
${\Delta}T_2=3$, and $\gamma_1=\gamma_2=1$.}
\label{fig:diagram}
\end{figure}

\begin{figure}
\centering
\includegraphics[width=\linewidth]{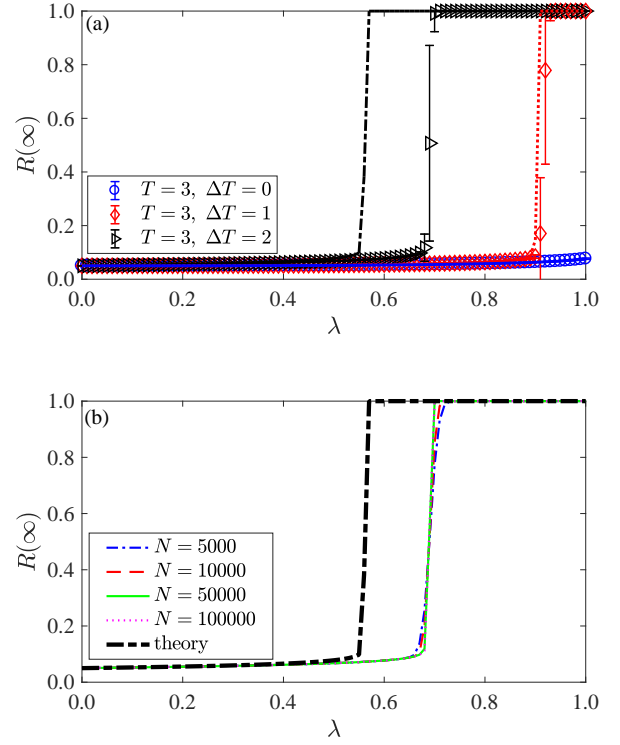}
\caption{ {\bf Behavioral adoption dynamics under symmetrical synergistic
interactions}. For random regular double-layer networks, (a) the fraction of
recovered nodes $R(\infty)$ [i.e., $R_1(\infty)=R_2(\infty)\equiv R(\infty)$]
versus $\lambda$, where $\lambda_1=\lambda_2\equiv\lambda$. The symbols are simulation results and the solid lines are the theoretical prediction obtained by iterating Eqs.~(\ref{eq:e24}) and (\ref{eq:e27}). (b) The simulation results of $R(\infty)$ versus $\lambda$ when $T=3$ and $\Delta T=2$ for different network sizes $N$. Other parameters are
$\gamma_1=\gamma_2=1$.}
\label{fig:symmetry}
\end{figure}

Our analysis and numerical computations indicate that, with synergistic
interactions between the spreading dynamics of two behaviors, both
$\lambda_1$ and $\lambda_2$ can affect $R_2(\infty)$ and the associated
phase transition characteristically. To further demonstrate the role of the
synergistic interactions, we show in Fig.~\ref{fig:diagram} color coded
values of $R_2(\infty)$ in the parameter plane ($\lambda_1$, $\lambda_2$)
for $T_1=1$, $T_2=4$, ${\Delta}T_1=0$, and ${\Delta}T_2=3$. There are three
regions in the parameter plane, determined by the two vertical lines at
$\lambda_1^{m}$ and $\lambda_1^{c}$, respectively, which are associated with
characteristically distinct behavioral adoption dynamics. In region I
($\lambda_1<\lambda_1^{m}$), only a small fraction of the nodes in layer $b$
adopt behavior 2. In region II ($\lambda_1^{m}<\lambda_1\leq\lambda_1^{c}$),
there is a discontinuous phase transition, where a larger fraction of
nodes adopt behavior $2$ for $\lambda_2>\lambda_2^c$ (white solid line).
In region III ($\lambda_1>\lambda_1^{c}$), there is a continuous phase
transition. The distinct types of phase transition are predicted through
our bifurcation analysis in Sec.~\ref{sec:theory}.

To gain further insights into the effects of synergistic interactions in
behavioral adoption dynamics, we study the special case where the two types
of behaviors are completely symmetric to each other. Fig.~\ref{fig:symmetry} (a)
shows, for $T_1=T_2=T$, ${\Delta}T_1={\Delta}T_2 \equiv \Delta{T}$, and
$\lambda_1=\lambda_2 \equiv \lambda$, the dependence of $R(\infty)$ on
$\lambda$ for different values of $\Delta{T}$. In the absence of synergistic
interactions, i.e., when the adoptions of behaviors $1$ and $2$ have no
effect on each other, neither behavior can spread globally and either
behavior can only be adopted by a small fraction of the nodes in the network.
For $\Delta{T}>0$ (i.e., $\Delta{T}=1,2$), the nodes that have adopted
behavior $1$ ($2$) only need additional $T-\Delta{T}$ pieces of information
to adopt behavior $2$ ($1$). As a result, the mutually cooperative spreading
of behaviors $1$ and $2$ leads to a wide adoption of both behaviors.
Increasing the synergistic interaction strength makes the dynamical
correlation between the two layers stronger. The discontinuous phase is more clear when the network size is enlarged. However, the improvement in decreasing the deviation of the critical threshold is less, as shown in Fig. 6 (b). In this regime, the deviation is mainly because the theoretical method can not capture the strong dynamical correlation between layers.

\begin{figure}
\centering
\includegraphics[width=\linewidth]{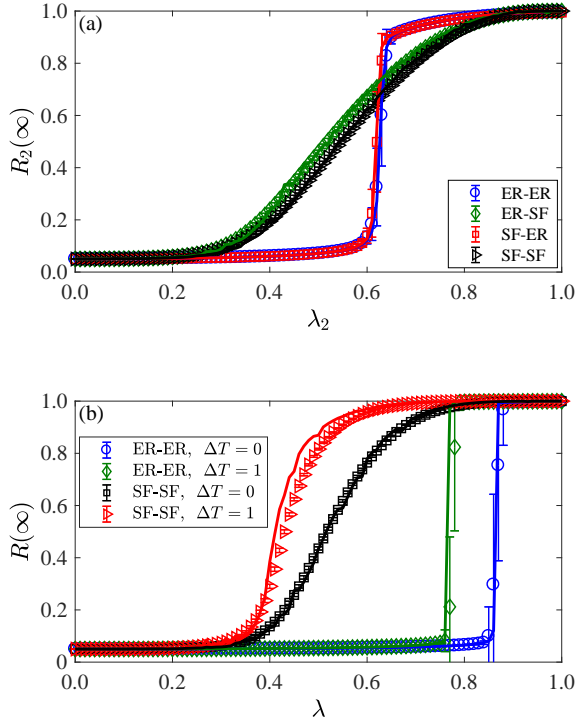}
\caption{ {\bf Synergistic behavior spreading on a multiplex networked
system with heterogeneous network layers}. For $T_1<T_2$, (a) $R_2(\infty)$
versus $\lambda_2$, where $T_1=1$, $T_2=4$, $\Delta{T_1}=0$, and
$\Delta{T_2}=2$. (b) The fraction of recovered nodes $R(\infty)$ versus
$\lambda$. The parameters are $T_1=T_2=3$, ${\Delta}T_1={\Delta}T_2=\Delta{T}$,
$\lambda_1=\lambda_2=\lambda$ and $R_1(\infty)=R_2(\infty)=R(\infty)$. The
symbols are simulation results and the solid lines are theoretical
prediction. Other parameters are $\gamma_1=\gamma_2=1$.}
\label{fig:SF}
%\end{center}
\end{figure}

\subsection{General multiplex networks}

We consider more general network topology for the network layers
in the multiplex system, such as ER-ER, SF-SF, ER-SF and SF-ER.
We use the standard configuration model~\cite{CBP:2005} to construct SF
networks with the degree distribution $P(k)={\Gamma}k^{-\gamma}$, where
$\gamma = 3$ is the degree exponent and the coefficient is
$\Gamma=1/\sum_{k_{min}}^{k_{max}}k^{-\gamma}$ with the minimum degree
$k_{min}=3$ and maximum degree $k_{max}{\sim}N^{1/(\gamma-1)}$. The
average degrees of SF and ER networks are set as $\langle k \rangle=10$,
and the network size is $N=5\times10^4$. For $T_1<T_2$, e.g., $T_1=1$ and
$T_2=4$, we fix the final adoption size of behavior 1 and vary the type of
network in layer $a$.

To facilitate comparison, we set $\lambda_1 = 0.12$
when layer $a$ is an ER network and $\lambda_1 = 0.113$ if network $a$ is
SF, so that the final adoption sizes of behavior 1 for both cases are
approximately $0.44$. As shown in Fig.~\ref{fig:SF}(a), the network type
in layer $a$ over which behavior 1 spreads has little effect on the spread
of behavior 2. For the symmetric case $T_1=T_2$, the dependence of $R(\infty)$
on $\lambda$ changes from being discontinuous to continuous as the network
becomes more heterogeneous (i.e., SF)~\cite{WTZL:2015}, as a strong
heterogeneity makes it harder for nodes in the subcritical state to adopt
a behavior simultaneously. Regardless of the network type, in general
synergistic interactions can facilitate adoption of both behaviors and
alter the nature of the associated phase transition.

\section{Discussion} \label{sec:discussion}

To understand social contagions in the human society at a quantitative
level is of great importance in the modern time. While the spread of a
single contagion can be analyzed through the traditional models of network
spreading dynamics, the simultaneous presence and spreading of two or
more contagions poses a challenge due to the mutual interplay between the
underlying dynamical processes. As an initial effort to address this problem,
we articulate a spreading model of multiple social behaviors on multiplex
networks subject to synergistic interactions. For simplicity, we consider
two-layer coupled networks and limit the number of distinct behaviors to
two: one on each layer. The manifestation of the synergistic mechanism is
that the adoption of the behavior by a node in one layer will increase the
chance for the node that is simultaneously present in the other layer to adopt
the behavior that spreads in that layer. The concrete setting enables us to
develop an edge-based compartmental theory and a bifurcation analysis to
uncover and explain how the synergistic interactions affects the spreading
dynamics in terms of the final adoption size and the distinct phase
transitions.

There are two types of synergistic interactions: asymmetric and symmetric.
In the asymmetric case, the adoption threshold of one behavior in one network
layer is less than that of the other behavior in the other layer. In this
case, the adoption of the behavior with the higher threshold has no effect
on the adoption of the other behavior. However, synergistic interactions
can promote the adoption of both behaviors. In fact, the interaction strength
and the information transmission rate of the behavior with the smaller
threshold value can affect the nature of the phase transition of the
behavior with the larger threshold: a small (large) value of the transmission
rate of the former can lead to a discontinuous (continuous),
first-(second-) order phase transition in the latter. In addition, a two
stage spreading process arises: nodes adopting the small threshold behavior
in one layer are more likely to adopt the large threshold behavior in the
other layer, which stimulates the remaining nodes in this layer to quickly
adopt the behavior. In the case of symmetric synergistic interactions, the
adoption processes in both layers can affect each other on an equal footing.
In this case, the interactions will greatly enhance the spreading of both
behaviors in their respective layers through a first-order phase transition.

Many issues remain, such as the effect of heterogeneity in the
synergistic strengths of the individual nodes on behavioral spreading and
the impacts of degree correlation between the network layers. 
In general, there are two kinds of dynamical correlation: intralayer
and interlayer. In each layer, the correlation can be described
by the edge-based compartmental theory. To make a theoretical
analysis feasible, we have neglected interlayer correlation, i.e., the
dynamical correlation among nodes in distinct layers.
However, in real situations, dynamical correlation
may exist between the same node in different layers, depending on the
strength of the synergistic interaction. If the interaction
strength is not too large, interlayer dynamical correlation is weak.
In this case, there is a good agreement between the theoretical prediction
and the simulation results (e.g., Figs.~\ref{fig:RR_RR} and 
\ref{fig:deltaEffect}). For relatively strong synergistic interaction 
(e.g., Fig.~\ref{fig:symmetry} for $\Delta T=2$), the simulation
results deviate from the theoretical prediction. Increasing the size of network will not help reduce the deviation, as interlayer
correlation can no longer be regarded as insignificant. A more accurate 
theory incorporating interlayer correlation is thus needed for 
synergistic affected information spreading in the strong interaction 
regime~\cite{WTSB:2017}.

\section*{Acknowledgements}
We are very grateful for the comments of anonymous reviewers. 
We thank A.~Vespignani and Q.  Zhang at the Laboratory for Modeling of 
Biological Socio-technical Systems (MOBS LAB) for valuable discussions and 
comments. This work was supported by the National Natural Science Foundation 
of China under Grants Nos.~11575041 and 61673086, the program of China 
Scholarships Council (No.~201606070059), and the Fundamental Research Funds 
for the Central Universities (Grant No.~ZYGX2015J153). YCL would like
to acknowledge support from the Vannevar Bush Faculty Fellowship program
sponsored by the Basic Research Office of the Assistant Secretary of Defense
for Research and Engineering and funded by the Office of Naval Research
through Grant No.~N00014-16-1-2828.

\bibliographystyle{apsrev4-1}
\bibliography{Multiple_Behavior}

\end{document}